# Heat Transfer and Cooling Techniques at Low Temperature

*B. Baudouy[1]*
CEA Saclay, France

**Abstract**
The first part of this chapter gives an introduction to heat transfer and cooling techniques at low temperature. We review the fundamental laws of heat transfer (conduction, convection and radiation) and give useful data specific to cryogenic conditions (thermal contact resistance, total emissivity of materials and heat transfer correlation in forced or boiling flow for example) used in the design of cooling systems. In the second part, we review the main cooling techniques at low temperature, with or without cryogen, from the simplest ones (bath cooling) to the ones involving the use of cryocoolers without forgetting the cooling flow techniques.

*Keywords*: heat-transfer, cooling techniques, low-temperature, cryogen.

## 1 Introduction

Maintaining a system at a temperature much lower than room temperature implies that the design of your cooling system must ensure its thermal stability in the steady-state regime; that is, the temperature of your system must remain constant in the nominal working condition. It also requires a certain level of thermal protection against transient events; that is, your temperature system must stay below a certain value to avoid problematic situations such as extra mechanical constraints due to the thermal expansion of materials with temperature. The ultimate goal is to minimize the heat load from the surroundings and to maximize the heat transfer with a cooling device.

To be able to achieve this objective, identification of the different heat loads on the system is required, as well as knowledge of the fundamental laws of heat transfer, the thermo-physical properties of materials and fluids such as the density (kg·m$^{-3}$) and heat capacity (J·kg$^{-1}$·K$^{-1}$), and thermal conductivity (W·m$^{-1}$·K$^{-1}$) and the cooling techniques such as conduction, radiation or forced flow based techniques.

The objective of the present chapter is to give an introduction to heat transfer and cooling techniques at low temperature, exemplified with practical cases and data.

## 2 Heat transfer at low temperature

### 2.1 Thermal conduction

#### 2.1.1 *Conduction in solids*

Thermal conduction is a heat transfer without mass transfer in solids. The relationship between the heat flux density and the temperature gradient, as presented in Fig. 1, is given by the Fourier law,

$$\vec{q} = -k(T)\vec{\nabla}T, \qquad (1)$$

---

[1] bertrand.baudouy@cea.fr

where $q$ is the heat flux density (W·m$^{-2}$), $k$ is the thermal conductivity (W·m$^{-1}$·K$^{-1}$), and $T$ is the temperature. In one dimension, Eq. (1) can be written as

$$q = -k(T)\frac{dT}{dx} \Rightarrow Q\int_0^L \frac{dx}{A} = \int_{T_{cold}}^{T_{hot}} k(T)dT, \qquad (2)$$

where $A$ is the cross-section of the domain and $Q$ is the power (W). In the case where $A$ is constant, then Eq. (2) is simplified to

$$q = -k(T)\frac{dT}{dx} \Rightarrow \frac{Q}{A} = \frac{1}{L}\int_{T_{cold}}^{T_{hot}} k(T)dT. \qquad (3)$$

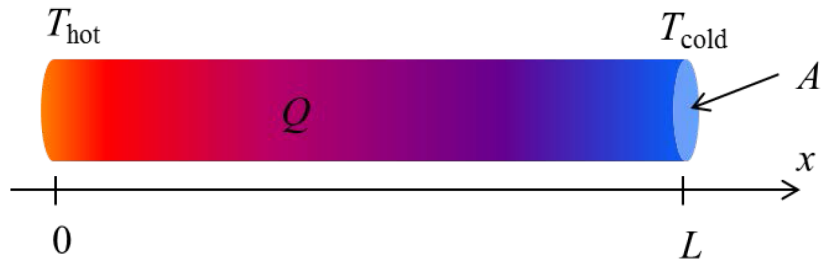

**Fig. 1:** The definition of heat flux in a one-dimensional domain

$\int_{T_{cold}}^{T_{hot}} k(T)dT$ is the thermal conductivity integral and knowledge of this property is of great importance in the thermal design of low-temperature devices, because the thermal conductivity of most materials varies strongly with temperature. Figure 2 depicts the evolution of the thermal conductivity integral with temperature for the common materials used at low temperature [1]. As expected, the thermal conductors possess a larger thermal conductivity integral.

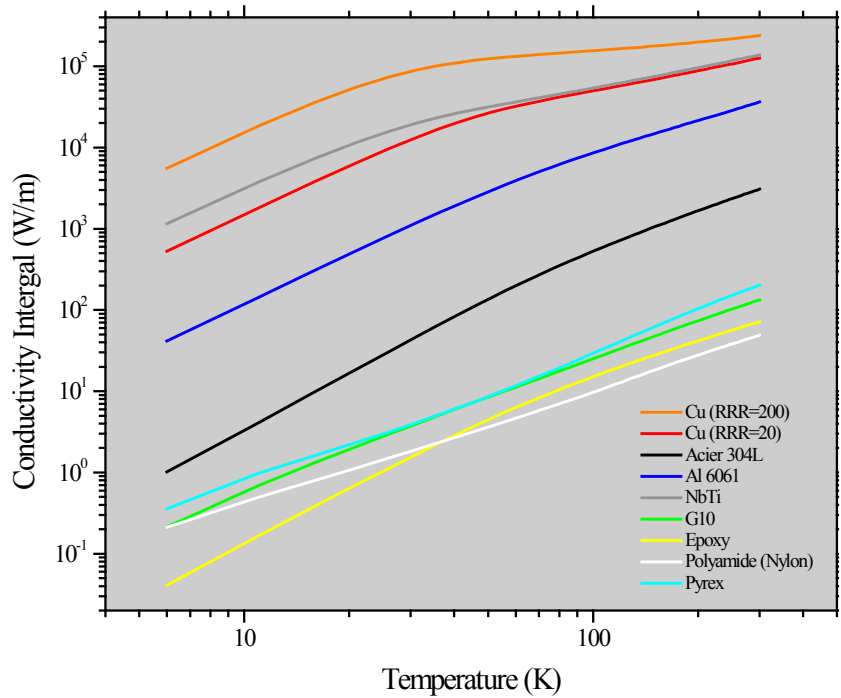

**Fig. 2:** The thermal conductivity integral of commonly used materials at low temperature [1]

One of the principal uses for the thermal conductivity integral is in the determination of heat losses and heat interception between room temperature and the low temperature of the system under consideration. A typical example is the computation of the heat input into a liquid helium bath cooled system, which is suspended by three 304 stainless steel rods from the 300 K top flange (cf. Fig. 3(a)). If the rods are 1 m long and 10 mm in diameter, then according to [1], the heat leak is

$$Q_{4K} = \frac{A}{L} \int_{4.2}^{300} k_{SS}(T)\,dT = \frac{2.36 \times 10^{-4}}{1} 3.07 \times 10^{3} = 0.7 \text{ W}. \tag{4}$$

This corresponds to a consumption of 1 l·h$^{-1}$ of liquid helium. Note that if the rods are made of copper (Residual Resistivity Ratio, RRR = 20) with a conductivity integral of 1.26 10$^5$ W·m$^{-1}$, then $Q_{4K} \approx 20$ W; or alternatively, in G10 (epoxy fiberglass tape) with an integral of 167 W·m$^{-1}$, then $Q_{4K} \approx 26$ mW. One has to keep in mind that the choice of the material also depends on other criteria, such as the mechanical, magnetic, and environmental conditions, and so on.

To reduce the heat load on the helium bath, heat interception by another cold source at an intermediate constant temperature (thermalization) is usually used. Historically, the most commonly used method is a heat sink cooled by boiling nitrogen; possibly replaced in recent years by a heat sink temperature-regulated by a cold stage of a cryocooler. In the case of boiling nitrogen (at 77 K) with an interception located at one third of the length from the top of the cryostat (see Fig. 3(b)), the heat leak to the liquid helium is reduced to

$$Q_{4K} = \frac{A}{L} \int_{4.2}^{77} k(T)\,dT = \frac{2.36 \times 10^{-4}}{0.75} 325 = 0.1 \text{ W}, \tag{5}$$

which corresponds to a seven-fold reduction of liquid helium consumption. The heat arriving at the nitrogen reservoir is computed as follows:

$$Q_{77K} = \frac{A}{L} \int_{77}^{300} k(T)\,dT = \frac{2.36 \times 10^{-4}}{0.25} 2.75 \times 10^{3} = 2.6 \text{ W}, \tag{6}$$

which is equivalent to a liquid nitrogen consumption of 0.06 l·h$^{-1}$. This example is simplified, but it shows the principle of heat interception. The optimization of the heat interception depends on many parameters, such as the thermalization temperature, the material properties, the geometry of the system, and so on.

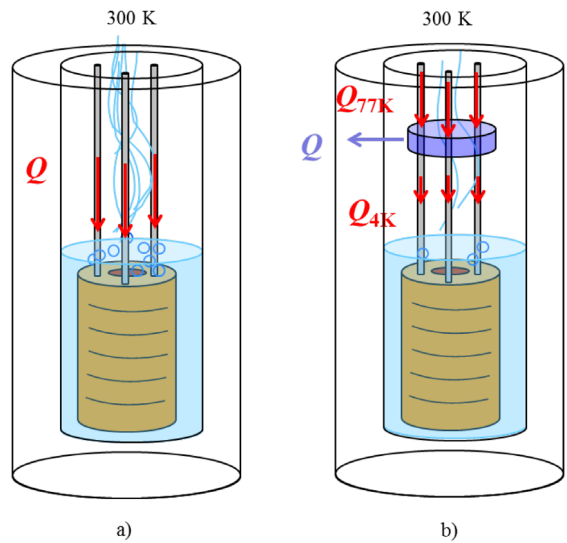

**Fig. 3:** The heat interception concept at 77 K

The thermal design at low temperature requires evaluation of the thermal losses of the system. To do so, one must calculate the thermal resistance of the structural material. In the steady-state regime and without internal dissipation, a thermal resistance $R_{th}$ can be defined from Eq. (3) as

$$Q \int_{Z_1}^{Z_2} \frac{dz}{A(z)} = -\int_{T_1}^{T_2} k(T) dT = \bar{k}(T_1 - T_2) \Rightarrow R_{th} = \frac{T_1 - T_2}{Q} = \frac{1}{\bar{k}} \int_{Z_1}^{Z_2} \frac{dz}{A(z)} \left[\frac{K}{W}\right], \quad (7)$$

where $\bar{k}$ is the average conductivity over the temperature range considered and $A(z)$ is the cross-section as shown in Fig. 4. For simple configurations, the resistance can be defined easily, as for a slab with a constant section ($A = A_1 = A_2$), $R_{th} = L/\bar{k}A$ (for a cylindrical wall between two radii, $R_1$ and $R_2$, with a length $L$, $R_{th} = \ln(R_2/R_1)/\bar{k}\, 2\pi L$). In the same manner, a convective boundary thermal resistance can also be simulated by considering a heat transfer coefficient $h$ and a heat transfer area $A$ as $R_{th} = 1/hA$.

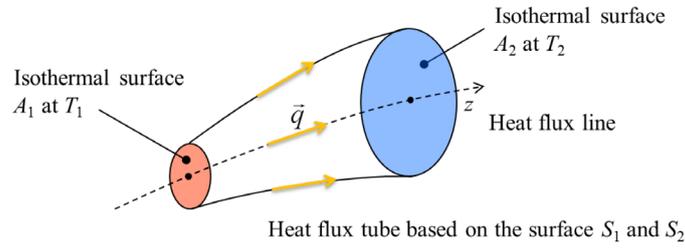

**Fig. 4:** A heat flux tube based on the surfaces $A_1$ and $A_2$

When several materials enter into the design of the structural component, it is necessary to account for all of them. When the components are thermally in series, then different values of $R_{th}$ are in series, so $R_{total}$ is simply $\sum R_i$. For the case of a composite wall with different heat transfer coefficients at the boundaries, as described in Fig. 5(a), the heat flux is computed as follows:

$$Q = \frac{T_h - T_c}{\sum_i R_i} = A \frac{T_h - T_c}{(1/h_h) + (L_1/\bar{k}_1) + (L_2/\bar{k}_2) + (L_3/\bar{k}_3) + (1/h_c)}. \quad (8)$$

In the case of parallel components, the $R_{th}$ are in parallel, so $1/R_{total} = \sum 1/R_i$. Then, the heat flux is as follows:

$$Q = A \frac{T_h - T_c}{(1/h_h) + (L_1/\bar{k}_1) + (2L_2/(\bar{k}_2 + \bar{k}_3)) + (L_4/\bar{k}_4) + (1/h_c)}. \quad (9)$$

Here, $L_2 = L_3$ and $A_2 = A_3 = A/2$.

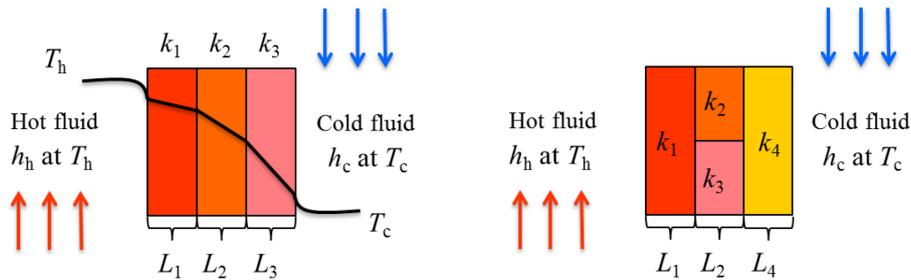

**Fig. 5:** 'Parallel' (a) and 'series/parallel' (b) composite walls with heat transfer coefficients at the boundaries

In the treatment of the thermal resistance, we assumed a perfect thermal contact between the different components. But this is rarely the case, since the two surfaces of the materials are not in a perfect and full contact. The rugosity creates an imperfect contact, generating a temperature drop (cf. Fig. 6) due to the local contact creating constriction of the flux lines. Additionally, the phonon scattering at the solid–solid contact (Kapitza resistance) and the heat transfer via eventual interstitial elements increase the contact thermal resistance significantly. As before, this resistance is defined as $R_c = (T_2 - T_1)/Q$. Resistance $R_c$ depends on the surface condition, the nature of the materials, the temperature, the interstitial materials, the compression force, and so on. It is proportional to the force applied on the contact, but not to the pressure (the number of contact points increases with the force). It reduces with increasing force but increases with decreasing temperature by several orders of magnitude from 200 to 20 K. At low temperature, $R_c$ can be the largest thermal resistance source, but it can be reduced by firm tightening of the two pieces or the insertion of conductive and malleable fillers (charged grease, indium, or surface coatings). Finally, it is worth saying that thermal contact resistance modelling is very difficult; therefore the use of experimental data is recommended, as in the example presented in Fig. 7 [2], where the thermal conductance values (i.e. the inverse of $R_c$) are presented.

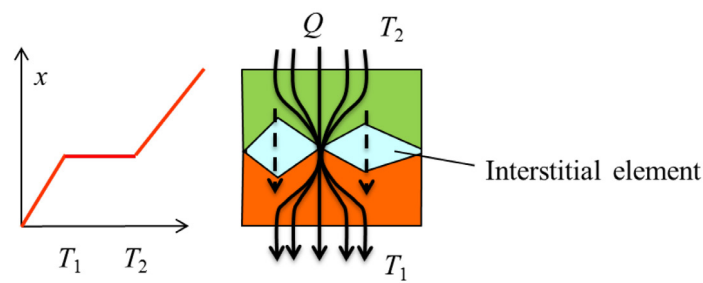

**Fig. 6:** The concept of contact thermal resistance

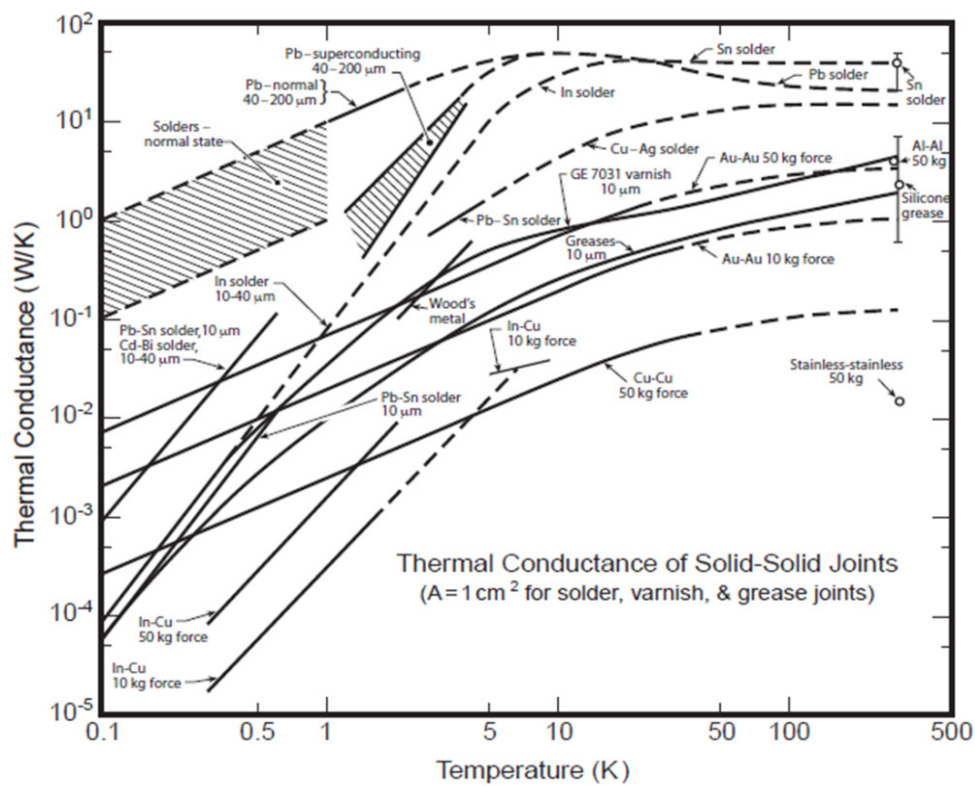

**Fig. 7:** The thermal conductance of different solid–solid joints as a function of temperature. Reproduced from [2] with permission of Oxford University Press.

It is often important to include the transient conduction process in a cryostat design at low temperature. In thermal conduction, transient processes are approached using an energy conservation equation that leads to a diffusion equation of the following type:

$$\rho C \frac{\partial T}{\partial t} = \nabla \cdot \left( -k(T) \vec{\nabla} T \right) + Q \left[ \text{W} \cdot \text{m}^{-3} \right], \quad (10)$$

where $\rho$ is the density and $C$ is the specific heat of the solid. On the left-hand side of the equation, the term accounts for the thermal inertia. The first term on the right-hand side accounts for the conduction, and the second one, $Q$, is a heat source. In 1D with constant thermal properties, Eq. (10) is simplified to

$$\frac{\partial T}{\partial t} = \frac{k}{\rho C} \frac{\partial^2 T}{\partial x^2} + Q^* \Rightarrow D = \frac{k}{\rho C} \left[ \text{m}^2 \cdot \text{s}^{-1} \right], \quad (11)$$

in which the thermal diffusivity $D = k/\rho C$ is defined. $D$ allows the evaluation of the characteristic diffusion time $\tau$ due to a thermal perturbation in the solid medium. This thermal time constant is obtained by solving Eq. (11) and is defined as

$$\tau \approx \frac{4}{\pi^2} \frac{L^2}{D}, \quad (12)$$

where $L$ is the characteristic thermal length of the solid. Equation (12) is the time constant when the temperature has reached two thirds of the final temperature. To reach 95% of the final temperature, the thermal time constant is $3\tau$. As can be seen from the equation, $\tau$ is a function of the length and of the thermal diffusivity and it differs from one material to another (Table 1). One can note the large increase of the thermal diffusivity with temperature.

Table 1: The thermal diffusivity of different materials at different temperatures in cm$^2 \cdot$s$^{-1}$ [1]

|  | 300 K | 77 K | 4 K |
| --- | --- | --- | --- |
| Cu OFHC (RRR = 150) | 1.2 | 3.2 | 11 700 |
| Pure Al (RRR = 800) | 1 | 4.7 | 42 000 |
| Commercial Al (6061) | 0.7 | 1.3 | 1 200 |
| SS 304 L | 0.04 | 0.05 | 0.15 |
| Nb–Ti | 0.03 | 0.02 | 0.51 |

### 2.1.2 Conduction in liquids

Liquids are bad thermal conductors, at room temperature and at low temperature alike. Furthermore, their thermal conductivity usually decreases with temperature. Table 2 presents the thermal conductivity of some common cryogens at atmospheric pressure [3, 4]. In most heat transfer situations at low temperature, the thermal conduction in a liquid is considered negligible compared to convection or phase change phenomena. There is one exception to this rule; namely, the superfluid helium. This topic will be treated in another section.

Table 2: The thermal conductivity of some cryogens at atmospheric pressure (W·m$^{-1}$·K$^{-1}$)

| O$_2$ ($T$ = 90 K) | N$_2$ ($T$ = 77 K) | H$_2$ ($T$ = 20 K) | He ($T$ = 4.2 K) |
| --- | --- | --- | --- |
| **0.152** | 0.14 | 0.072 | 0.019 |

## 2.1.3 Conduction in gas

The heat transfer between two surfaces in a gas is of interest to evaluate the heat leak or to characterize thermal switches, devices that exchange heat with gas in certain conditions. There are two different heat transfer regimes depending on the ratio between the mean free path of the gas molecules, $\lambda$, and the distance $L$ between the two surfaces involved in the heat transfer:

$$\lambda = \frac{RT}{\sqrt{2}\pi d^2 N_A p} \quad \begin{cases} R = \text{universal gas constant} = 8.31 \text{ J} \cdot \text{mol}^{-1} \cdot \text{K}^{-1}, \\ N_A = \text{Avogadro's number} = (6.02 \times 10^{23}) \cdot \text{mol}^{-1}, \\ d = \text{molecule diameter}. \end{cases} \quad (13)$$

When $\lambda \gg L$, it corresponds to the *free molecular regime*, whereas when $\lambda \ll L$, it corresponds to the *hydrodynamic regime*. The free molecular regime is obtained at low residual pressure, and the heat transfer depends on the residual gas pressure and is independent of $L$. The heat leak, $Q$, in the free molecular regime is described by Kennard's law:

$$Q = A\alpha \left(\frac{\gamma+1}{\gamma-1}\right) \sqrt{\frac{R}{8\pi M}} \frac{\Delta T}{\sqrt{T}} p \quad \text{with} \quad \gamma = \frac{C_p}{C_v}, \quad (14)$$

where $M$ is the molar mass of the gas, $A$ is the surface area receiving the heat flux, and $\alpha$ is an accommodation coefficient [2]. Coefficient $\alpha$ relates the degree of thermal equilibrium between the gas and the wall. Its value ranges from $\alpha \leq 0.5$ for helium, to $\alpha \sim 0.78$ for argon and $\alpha \sim 0.78$ for nitrogen. The hydrodynamic regime is obtained at high residual gas pressure and the heat transfer is independent of pressure and described by a Fourier law. The thermal conductivity is thus derived from the kinetic theory. Some values at atmospheric pressure are presented in Table 3, and Fig. 8 presents a very useful example of heat transfer in helium gas between two copper plates separated by 1 cm [2].

**Table 3:** The thermal conductivity, $k$ [mW·m$^{-1}$·K$^{-1}$] at 1 atm

| T (K) | $^4$He | H$_2$ | N$_2$ |
|---|---|---|---|
| 300 | 150.7 | 176.9 | 25.8 |
| 75* | 62.4 | 51.6 | 7.23 |
| 20 | 25.9 | 15.7 | |
| 5 | 9.7 | | |

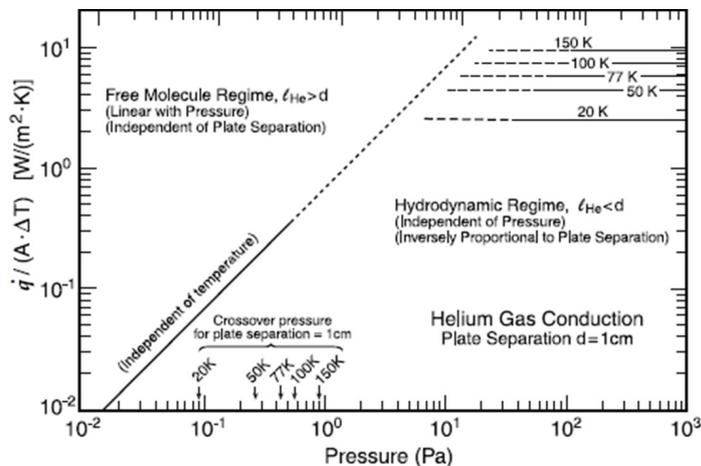

**Fig. 8:** The heat transfer in helium gas between two copper plates separated by 1 cm. Reproduced from [2] with permission of Oxford University Press.

## 2.2 Thermal radiation

### *2.2.1 General laws*

Any surface at finite temperature absorbs, reflects, and emits electromagnetic radiation. The wavelengths associated with thermal radiation span from 0.1 to 100 μm, as shown in Fig. 9.

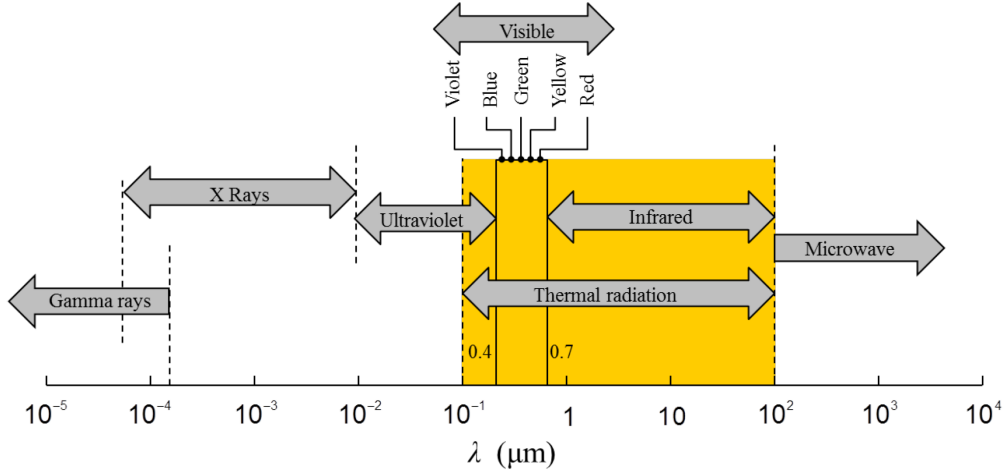

**Fig. 9:** The electromagnetic radiation spectrum, showing the thermal radiation zone in yellow

The emitted radiation, Φ, consists of a continuous non-uniform distribution of monochromatic components. This distribution and the magnitude of the emitted radiation depend on the nature and the temperature of the emitting surface. Φ also has a directional distribution (Fig. 10). Therefore to characterize the radiation heat transfer (as a function of temperature and surface), both the spectral and the directional dependence must be known. On a real body, incident radiation is absorbed, transmitted through the body or reflected; therefore the law of energy conservation requires $1 = \alpha + \tau + \rho$, where α represents the spectral absorption component, $\tau$ the spectral transmission component, and $\rho$ the spectral reflection component.

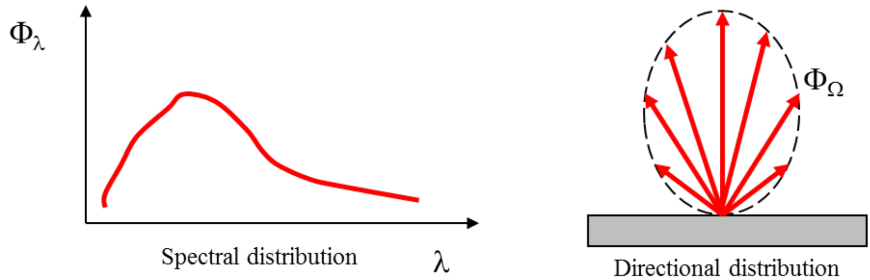

**Fig. 10:** Spectral and directional distribution

When describing the radiation characteristics of real surfaces, it is useful to start with the concept of an ideal surface: the blackbody. The blackbody is a perfect emitter and absorber. It absorbs all incident radiation regardless of the wavelength and direction. At a given temperature and wavelength, the emission is maximum and the blackbody is a diffuse emitter (no directional dependence). The emissive power (W·m$^{-3}$) of the blackbody per unit wavelength and per surface (hemispherical) as a function of wavelength is described by Planck's law:

$$E_\lambda^0 = \frac{C_1}{\lambda^5 \left(e^{C_2/\lambda T} - 1\right)} \quad \text{with} \quad \begin{cases} C_1 = 2\pi h C_0^2 = 3.742 \times 10^{-16} \text{ Wm}^4 \cdot \text{m}^{-2}, \\ C_2 = h C_0 / k = 1.4388 \times 10^{-16} \text{ Wm}^4 \cdot \text{m}^{-2}. \end{cases} \quad (15)$$

The emissive power is presented in Fig. 11 as a function of wavelength and temperature. It exhibits a maximum at a certain temperature, given by Wien's law:

$$\lambda_{max} T = 2898 \ \mu m \cdot K^{-1}. \tag{16}$$

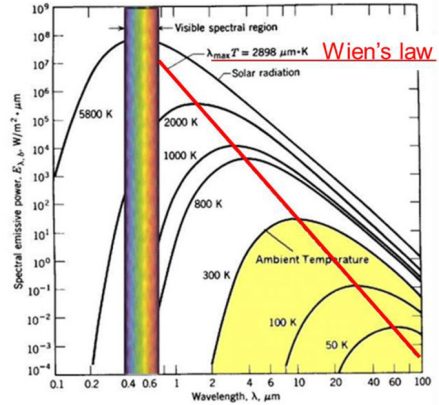

**Fig. 11:** The emissive power of a spectral blackbody

Integrating Planck's law with respect to wavelength, the so-called Stefan–Boltzmann law is obtained, defining the blackbody total hemispherical emissive power at temperature $T$:

$$E^0 = \int_0^\infty \frac{C_1}{\lambda^5 \left(e^{C_2/\lambda T} - 1\right)} d\lambda = \sigma T^4 \quad \text{with} \quad \sigma = 5.67 \times 10^{-8} \ W \cdot m^{-2} \cdot K^{-4}. \tag{17}$$

For a real surface material, the emissive power is only a fraction of the blackbody. The ratio of the real surface to the blackbody emissive power is called the emissivity, $\varepsilon$. One can define the spectral, monochromatic directional emissivity as $\varepsilon(\lambda, \theta, \phi, T)$ and the spectral emissivity as $\varepsilon(\lambda, T)$. But for simplicity, the total emissivity, $\varepsilon(T)$, is most commonly used. Then, the emissive power for a real surface material is defined as

$$E^0 = \varepsilon \sigma T^4. \tag{18}$$

In general, emissivity decreases with temperature but increases with oxidation, impurities, dirt, and so on. To achieve the lowest emissivity value, it is recommended that highly polished and highly conductive clean surfaces (e.g. gold, silver, copper, or aluminium) should be used. Table 4 presents the total emissivity of various materials used at cryogenic temperatures compared to 3M black paint at different temperatures [5]. Figure 12 also presents useful emissivity data for various materials between two surfaces at different temperatures [6]. These data help in the design of the thermal radiation shield, from room temperature down to liquid helium temperature.

**Table 4:** The total emissivity of various metals at three different temperatures

|  | 300 K | 78 K | 4.2 K |
|---|---|---|---|
| **3M black paint (80 μm) on a copper surface** | 0.94 | 0.91 | 0.89 |
| **Polished aluminium (33 μm roughness)** | 0.05 | 0.023 | 0.018 |
| **Polished copper (41 μm roughness)** | 0.10 | 0.07 | 0.05 |
| **304 Polished stainless steel (27 μm roughness)** | 0.17 | 0.13 | 0.08 |

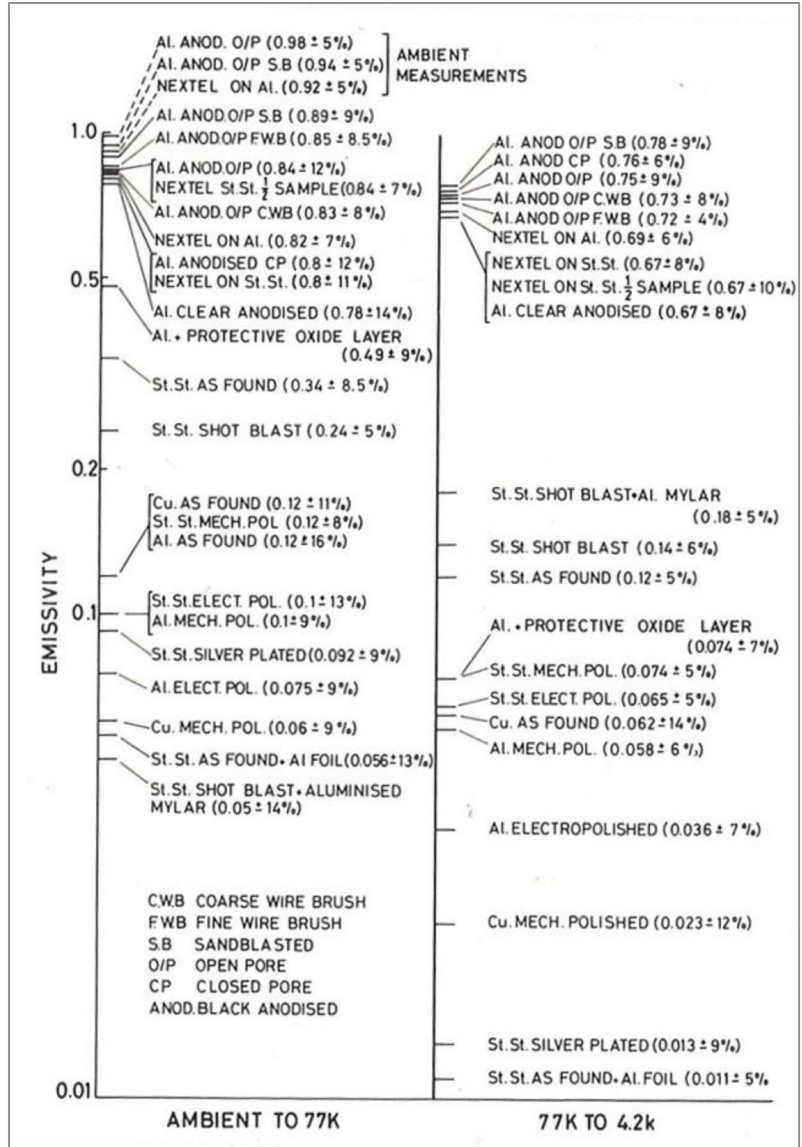

**Fig. 12:** The total emissivity from ambient temperature down to liquid helium temperature (4.2 K). Reproduced from [6] with permission of Springer.

*2.2.2 Radiation exchange between two surfaces*

The heat transfer by radiation between two enclosed surfaces; from one at temperature $T_1$ with an emissivity $\varepsilon_1$ and a surface $A_1$, to another at $T_2$ with $\varepsilon_2$ and $A_2$, can be written as follows:

$$q_{12} = \sigma(T_1^4 - T_2^4) / \left( \frac{1-\varepsilon_1}{\varepsilon_1 A_1} + \frac{1}{A_1 F_{12}} + \frac{1-\varepsilon_2}{\varepsilon_2 A_2} \right), \tag{19}$$

where $q_{12}$ is the heat transfer rate from surface 1 to surface 2. $F_{12}$ is the 'view factor', which is the fraction of the heat leaving surface 1 that is intercepted by surface 2. A useful relationship can be used to determine the view factors in complex geometry as the reciprocity relation arising from the definition of the view factor [7]:

$$A_i F_{ij} = A_j F_{ji}. \tag{20}$$

This relation is essential to determine one view factor from knowledge of the other. Another important view factor relation pertains to the enclosure surface. Again, from the definition of the view factor, one can show in this case that

$$\sum_{j=1}^{N} F_{ij} = 1. \quad (21)$$

In determining the heat balance in the apparatus design at low temperature, the most useful special examples of Eq. (19) are the large parallel plates,

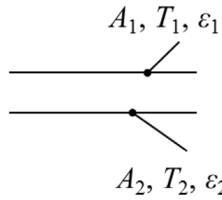

$$A = A_1 = A_2, \quad F_{12} = 1, \quad q_{12} = \frac{\sigma A_1 (T_1^4 - T_2^4)}{\frac{1}{\varepsilon_1} + \frac{1}{\varepsilon_2} - 1}; \quad (22)$$

and the long concentric cylinders,

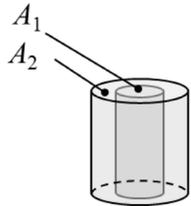

$$\frac{A_1}{A_2} = \frac{r_1}{r_2}, \quad F_{12} = 1, \quad q_{12} = \frac{\sigma A_1 (T_1^4 - T_2^4)}{\frac{1}{\varepsilon_1} + \frac{1-\varepsilon_2}{\varepsilon_2}\left(\frac{r_1}{r_2}\right)}. \quad (23)$$

### 2.2.3 Shielding and multilayer insulation

To understand the philosophy of 'passive' thermal shielding from room temperature to low temperature, simple evaluation using the blackbody relation $q = \sigma(T_{\text{warm}}^4 - T_{\text{cold}}^4)$ is sufficient, without knowledge of the emissivities and view factors. The heat transfer density from 300 K to 77 K using this equation is 457 W·m$^{-2}$ and from 300 K to 4.2 K, $q$ = 459 W·m$^{-2}$. In other words, from 77 K to 4.2 K, $q$ = 2 W·m$^{-2}$. The heat transfer rate from 77 K to 4.2 K is 200 times lower than that from 300 K to 77K (and 4.2 K). Obviously, an intermediate surface held at intermediate temperature is required to reduce the heat load at low temperature. Using Eq. (22) with the assumption that the surfaces and emissivities are identical for the different reflecting surfaces, one can calculate the heat flux density $q$ as a function of the number of intermediate surfaces for passive shielding, as shown in Table 5.

**Table 5:** Passive shielding heat transfer as a function of the intermediate surface number

| $q = \dfrac{\varepsilon\sigma}{2-\varepsilon}(T_{\text{warm}}^4 - T_{\text{cold}}^4)$ | $q = \dfrac{1}{2}\dfrac{\sigma\varepsilon}{2-\varepsilon}(T_{\text{warm}}^4 - T_{\text{cold}}^4)$ | $q = \dfrac{1}{n+1}\dfrac{\sigma\varepsilon}{2-\varepsilon}(T_{\text{warm}}^4 - T_{\text{cold}}^4)$ |
|---|---|---|
| 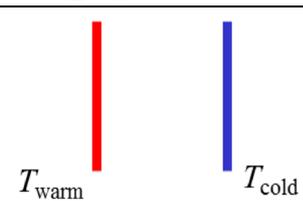 | 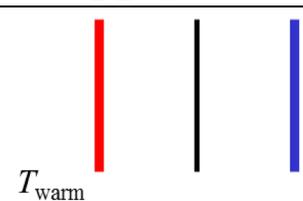 | 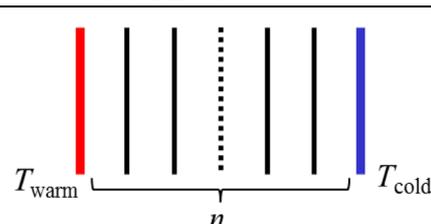 |
| | $T^4 = \dfrac{T_{\text{warm}}^4 + T_{\text{cold}}^4}{2}$ | $T_i^4 = T_{\text{cold}}^4 + \dfrac{T_{\text{warm}}^4 + T_{\text{cold}}^4}{i+1}$ |

In practice, passive thermal shielding is implemented using a multilayer insulation (MLI) system (also called superinsulation), an assembly of reflective films (usually aluminium or aluminized polyester film) separated by insulating interlayers (polyester, glass-fibre nets, or paper), operated under vacuum, as shown in Fig. 13. The reflecting layers reduce heat transfer by radiation, the insulating interlayers reduce heat transfer by conduction between reflecting layers, and the high vacuum reduces convection and residual gas conduction.

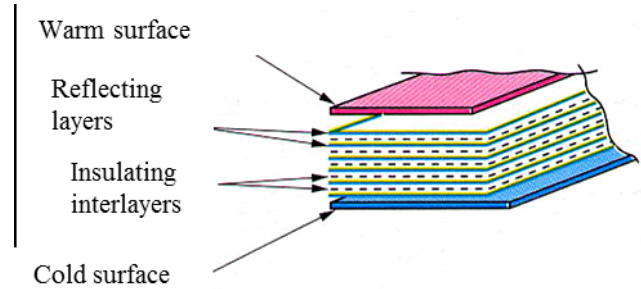

**Fig. 13:** A depiction of a multilayer insulation (MLI) system

Under a good vacuum, a typical value of the heat transfer for a 20-layer MLI system is around 1–3 W·m$^{-2}$ from 300 K to 80 K (but can reach 5 W·m$^{-2}$ if compressed) and less than 100 mW·m$^{-2}$ between 80 K and 4 K. To optimize the use of MLI, the maximum number of layers per centimetre should be between 20 and 30, and isothermal contact points are required. But the efficiency of the MLI can deteriorate due to bad installation; for example, gaps in the MLI blankets, overlapping, with the inner (cold) side in contact with the outer (hot) side, and mechanical stress. For comprehensive application data and empirical formulas, see Ref. [8].

## 2.3 Convection

### 2.3.1 Introduction

Once again, we will present the basic notion of the physics of convection to extract the useful information on cryogenic design. In convection heat transfer, the heat can be transferred in the fluid by movement of matter. It is a quantity of energy that is advected within the fluid. The movement of matter can be created externally by a pump or a pressurization system. Hence we talk about forced convection. In this regime, the equations of motion in the Boussinesq approximation are useful to present the dimensionless numbers and their physical significance. It will be also useful to evaluate heat transfer coefficients and flow regimes. In the steady-state regime, the equations of motion in the Boussinesq approximation are as follows:

$$
\begin{array}{ll}
\text{Continuity} & \nabla \cdot \mathbf{v} = 0 \\
\text{Navier–Stokes} & \rho \mathbf{v} \cdot \nabla \mathbf{v} = -\nabla p + \mu \nabla^2 \mathbf{v} + \mathbf{f} \\
\text{Energy} & \rho C \mathbf{v} \cdot \nabla T = k \nabla^2 T + Q
\end{array}
\quad \xrightarrow{\text{Dimensionless}} \quad
\begin{array}{l}
\nabla \cdot \mathbf{v}^* = 0 \\
\mathbf{v}^* \cdot \nabla \mathbf{v}^* = -\nabla p^* + \text{Re}^{-1} \nabla^2 \mathbf{v}^* \\
\mathbf{v}^* \cdot \nabla T^* = \text{Re}^{-1} \text{Pr}^{-1} \nabla^2 T^*
\end{array}
\quad (24)
$$

From the dimensionless equations, the flow is characterized by two important numbers: the Reynolds number, Re, which is the ratio of the inertia to the viscous forces,

$$\text{Re} = \frac{\rho L v}{\mu}, \quad (25)$$

and which characterizes the flow regime; and the Prandtl number, Pr, which is the ratio of the momentum to the thermal diffusivity,

$$\Pr = \frac{\mu C}{k},\qquad(26)$$

and which characterizes the properties of the fluid. It is essential to know the regime in which the flow is taking place, since the surface heat transfer and the friction both depend strongly on it. When Re < 2300, it is in the laminar regime, where the viscous forces dominate, creating an 'ordered' flow with streamlines. In this regime, the surface heat transfer is low and so is the surface friction. The turbulent regime is reached for $Re_L > 5 \times 10^5$, while $Re_D > 4000$ for a plate and a tube-shaped geometry, respectively. Here, the inertia forces dominate, and the flow becomes highly irregular (velocity fluctuation), as shown in Fig. 14 (velocity fluctuation). The surface heat transfer and the friction are higher than those in the laminar regime.

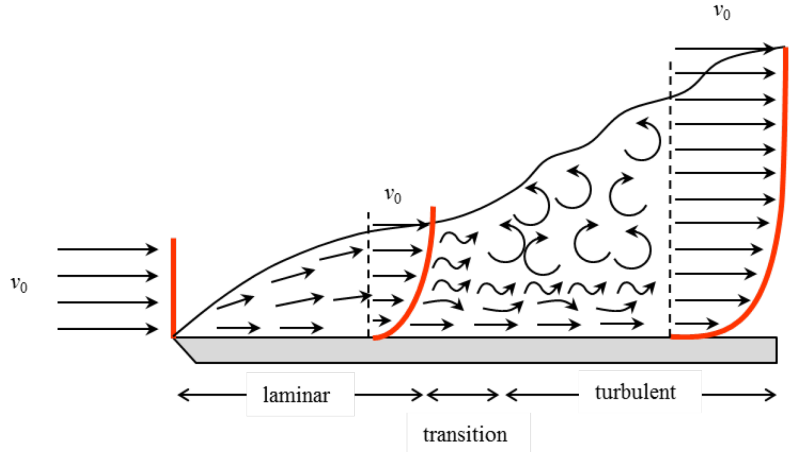

**Fig. 14:** The development of a velocity boundary layer in a fluid in contact with a plate

When the fluid movement is created internally, by a decrease or increase of the fluid density or by the buoyancy effect, it is called 'natural convection'. In the steady-state regime, the corresponding equations of motion in the Boussinesq approximation are as follows:

$$\begin{aligned}\nabla.\mathbf{v}&=0\\\rho\mathbf{v}.\nabla\mathbf{v}&=-\nabla p+\mu\nabla^2\mathbf{v}+\beta\Delta T\mathbf{g}\\\rho C\mathbf{v}.\nabla T&=k\nabla^2\end{aligned}\xrightarrow{\text{Dimensionless}}\begin{aligned}\nabla.\mathbf{v}^*&=0\\\mathbf{v}^*.\nabla\mathbf{v}^*&=-\nabla p^*+\text{Re}^{-1}\nabla^2\mathbf{v}^*+\text{Gr}\,\text{Re}^{-2}\,T^*\\\mathbf{v}^*.\nabla T^*&=\text{Re}^{-1}\Pr^{-1}\nabla^2 T^*\end{aligned}\qquad(27)$$

In natural convection, the main dimensionless number is the Grashof number,

$$\text{Gr}=\frac{g\beta\Delta T L^2}{\mu^2},\qquad(28)$$

which is the ratio of the buoyancy to the viscous forces. Looking at the dimensionless equation, one can deduce that when $\text{Gr} \times \text{Re}^{-2} \gg 1$, forced convection is negligible. If $\text{Gr} \times \text{Re}^{-2} \approx 1$, then the flow is in a mixed convection state, where natural and forced convections are equally important. The Grashof number has the same role in natural convection as Re in forced convection. Turbulence has a strong effect, as in forced convection, and is reached for $\text{Gr} \times \Pr \geq 10^5$.

When there is an interaction with solid surfaces in convection, the quantity of energy transferred in (or out) of the fluid to solids must be evaluated. The heat transferred to solid elements is modelled by Newton's law:

$$q=h(T_s-T_\infty),\qquad(29)$$

where $h$ is the heat transfer coefficient [W·m$^{-2}$·K$^{-1}$], $T_s$ the temperature of the solid and $T_\infty$ is the temperature far from the solid. At the boundary as depicted in Fig. 15, the local heat flux is $q_n = k.\nabla T_n$; that is, the heat flux going through the conductive layer of the fluid also obeys Newton's law. This representation leads to the dimensionless Nusselt number, defined as

$$\mathrm{Nu} = \frac{hL}{k} = \left.\frac{\partial T^*}{\partial n^*}\right|_0. \tag{30}$$

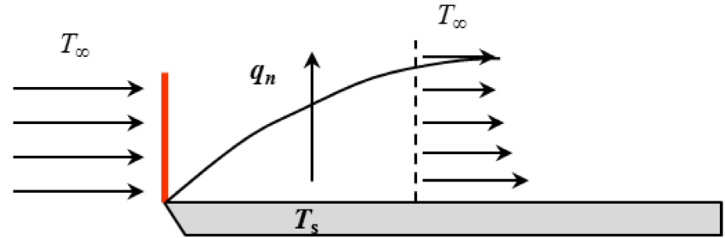

**Fig. 15:** The thermal boundary layer at the wall of a flow with a plate

The Nusselt number is to the thermal boundary what the friction coefficient is to the velocity boundary. The Nusselt number is defined as Nu = f(Re, Pr, $L$) for forced convection and Nu = f(Gr, Pr, $L$) for natural convection. Their form is different for a turbulent or a laminar regime, and in the following sections we will present the correlations that are useful for design at low temperature, depending on the regime and the configuration of the flow.

### 2.3.2  Natural convection

The heat flux at the wall is computed with correlations of the type Nu = f(Gr, Pr, $L$), where the thermophysical properties are usually established at an average temperature between the solid and the fluid temperatures. The heat transfer coefficients are of the order of 10–100 W·m$^{-2}$·K$^{-1}$. In natural convection, the simplest correlations are expressed as Nu = $c \times$ (Gr $\times$ Pr)$^n$, in which $n = 1/4$ for a laminar regime and $n = 1/3$ for a turbulent regime. The correlations are not very different from those for classical fluids, but few data exist for cryogenic fluids, since two-phase phenomena take over even at low heat fluxes. Table 6 shows various correlations taken from the literature [9-12].

**Table 6:** Nusselt correlations for cryogenic fluids

|  | $c$ | $n$ |
|---|---|---|
| **Turbulent supercritical helium (vertical orientation)** | 0.615 | 0.258 |
| **Turbulent liquid nitrogen (various orientations)** | 0.14 | 1/3 |
| **Turbulent liquid hydrogen (various configurations)** | 0.096 | 0.352 |

### 2.3.3  Forced convection

As for natural convection, the correlations used for non-cryogenic fluids are suitable at low temperature in the forced convection regime. For turbulent flows in pipes, the Dittus–Boetler correlation, Nu = 0.023Re$^{0.8}$Pr$^{0.4}$, is used for hydrogen [13]; a modified version, Nu = 0.022Re$^{0.8}$Pr$^{0.4}$, is used for supercritical helium [14]; and another modified version, Nu = 0.027Re$^{0.8}$Pr$^{0.14/3}(\mu_f/\mu_w)^{0.14}$, is used for nitrogen [15]. The heat transfer coefficients can go up to several kW·m$^{-2}$·K$^{-1}$. A laminar flow in pipes is very rare, except in porous media, and it is too specific for details to be given here.

## 2.3.4 Boiling convection

In boiling convection, heat is transferred between a surface and the fluid by the conjunction of a phase change and the vapour bubble movement in the vicinity of the surface. Heat transfer combines natural convection in the liquid, latent heat to be absorbed for the bubble formation, and the bubble hydrodynamics. The heat transfer process depends on the bubble growth rate, the detachment frequency, the number of nucleation sites, and the surface conditions. In pool boiling, several regimes can be identified, as shown in Fig. 16. Before the onset of boiling, natural convection takes place; and since the boiling heat transfer is extremely efficient, when boiling is activated, the wall temperature increase is slowed down. After the onset of boiling, the evolution of nucleate boiling is encountered; that is, from partially to fully developed nucleate boiling, where the vapour content and structure are continuously increasing. At one point, called the 'critical point', the vapour production is so high that the vapour structures coalesce and form a blanket of vapour at the heating surface. This results in a new regime called 'film boiling', where the heat is primarily transferred by conduction through the vapour film, explaining the large increase of the wall temperature. Table 7 shows the different heat flux and temperature increases at the critical points for various cryogenic fluids [10, 12, 16, 17].

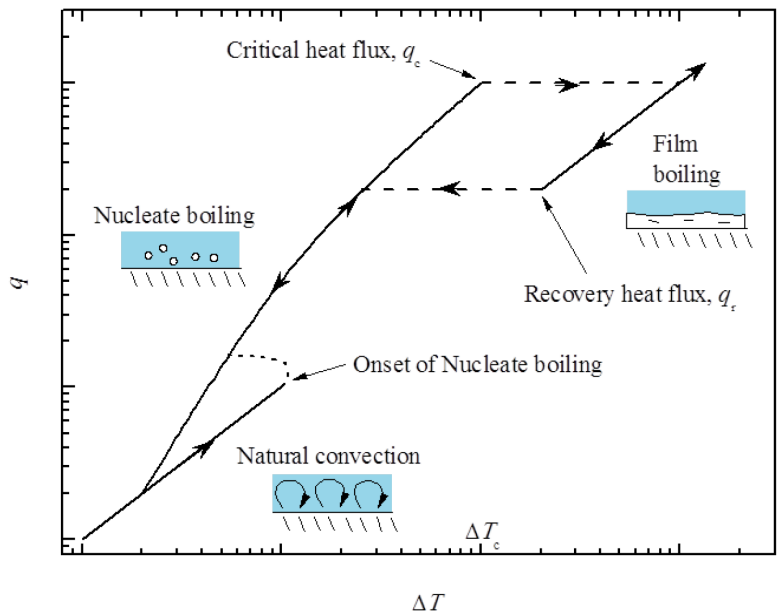

**Fig. 16:** The boiling regimes for a flat horizontal surface

**Table 7:** Characteristic points of the boiling curve

|  | $\Delta T_c$ (K) | $q_c$ (kW·m$^{-2}$) | $q_r$ (kW·m$^{-2}$) |
|---|---|---|---|
| **Helium** | 1 | 10 | |
| **Nitrogen** | 10 | 100 | |
| **Hydrogen** | 5 | 100 | 10 |

There are several heat transfer correlations that can fit experimental data within one order of magnitude and here we will present the most popular one – the Kutateladze correlation $q = f(p) \times \Delta T^{2.5}$, which works for most cryogenic fluids:

$$\frac{h}{k_l}\left(\frac{\sigma}{g\rho_\ell}\right)^{1/2} = 3.25 \times 10^{-4} \left[\frac{qC_{p\ell}\rho_\ell}{h_{\ell v}\rho_v k_\ell}\left(\frac{\sigma}{g\rho_\ell}\right)^{1/2}\right]^{0.6} \left[g\left(\frac{\rho_\ell}{\mu_\ell}\right)^2 \left(\frac{\sigma}{g\rho_\ell}\right)^{3/2}\right]^{0.125} \left(\frac{p}{(\sigma g\rho_\ell)^{1/2}}\right)^{3/2}, \quad (31)$$

where $\sigma$ is the tension surface and $p$ is the pressure. The subscripts $\ell$ and v stand for 'liquid' and 'vapour', respectively.

Figure 17 shows an example of experimental data for flat horizontal boiling heat transfer in nitrogen and the comparison with the Kutateladze correlation [10]. The critical heat flux can also be computed with Kutatekadze correlations, which once again give very good results compared to experimental data for several cryogenic fluids, such as helium, nitrogen, oxygen, and hydrogen [12, 18]. It is written as follows:

$$q_c = 0.16 h_{\ell v} \rho_v^{1/2} \left[ \sigma g (\rho_\ell - \rho_v) \right]^{1/4}, \tag{32}$$

where $h_{\ell v}$ is the latent heat of vaporization. It is worth noting that this correlation is not valid when the fluid is sub-cooled; that is, when the pressure above the heated surface is higher than the saturation pressure. For this peculiar case, the following correlation can be used [19]:

$$\frac{q_{c,\text{sub}}}{q_{c,\text{sat}}} = 0.2 \left[ 1 + 0.15 \left( \frac{\rho_\ell}{\rho_v} \right)^{3/4} \frac{C_p \Delta T_{\text{sub}}}{h_{\ell v}} \sigma \right]. \tag{33}$$

Last but not least, two-phase forced flow heat transfer modelling must take in account the boiling heat transfer, which depends on the surface heat transfer, and the forced convection, which depends on the vapour quality ($x = \dot{m}_v/\dot{m}_t$) and the total mass flow rate ($\dot{m}_t$). Usually, boiling tends to be dominant for low vapour quality and high heat flux, while forced convection tends to be dominant for high vapour quality and a large mass flow rate. Several general correlations specific to cryogenic fluids emerge. The best approach is to try more than one such correlation to evaluate the heat transfer rate. The most successful ones are the superposition method (Chen) [20], the intensification model (Shah) [21], and the asymptotic model (Liu and Winterton, $n = 2$) or the Steiner–Taborek ($n = 3$) correlation. The latter is considered by some authors as the most accurate and it includes the cryogenic fluids (helium, hydrogen, nitrogen, oxygen, etc.) [22].

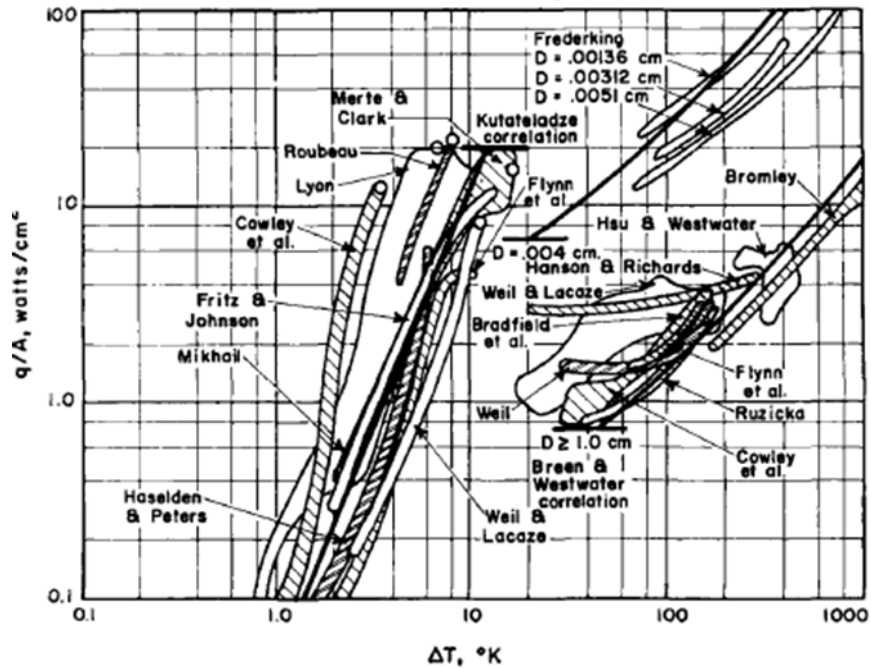

**Fig. 17:** Boiling curves for horizontal pool boiling for nitrogen, reprinted from [10] with permission of Elsevier

# 3 Cooling techniques at low temperature

## 3.1 Introduction

Nowadays, cooling a device to low temperature can be achieved with or without cryogenic fluids. A cooling method is called a 'wet method' when a cryogenic fluid is used in contact with the device or a 'dry method' when a cryocooler is used without any fluid as coolant. A third technique is the 'indirect cooling method', when a cryogenic fluid is used without direct contact with the device but only through intermediate components. Figure 18 presents a good summary of all three methods and their variants that are used to reach the low temperature.

The 'dry system' method – or simply the 'cryogen-free cooling method' – relies solely on conduction to cool down a system. The cold source in this method is a cryocooler. The main reason for using such a method is to avoid dealing with a cryogenic fluid. Due to the performance limitations of the cryocoolers available today, however, they are only suitable for systems subjected to a small heat load (typically, a few watts at 4 K) and with slow transient perturbation processes. Typical examples are room-temperature bore magnets or MRI magnets.

In the 'indirect cooling' method, there is no direct contact between the cryogen and the system; yet the heat conduction in the system is still of importance, but one now has to consider the surface heat transfer between the fluid and the cooling loop on the system. This method can be generally achieved with an external flow of cryogen as a cold source. The main reason for using this method is the reduction of the cryogenic fluid inventory while keeping a high fluid–solid heat transfer coefficient. It is suitable for moderate or well-distributed heat load systems and with slow transient perturbation processes. Typical examples are the large detector magnets, such as CMS or Atlas at CERN.

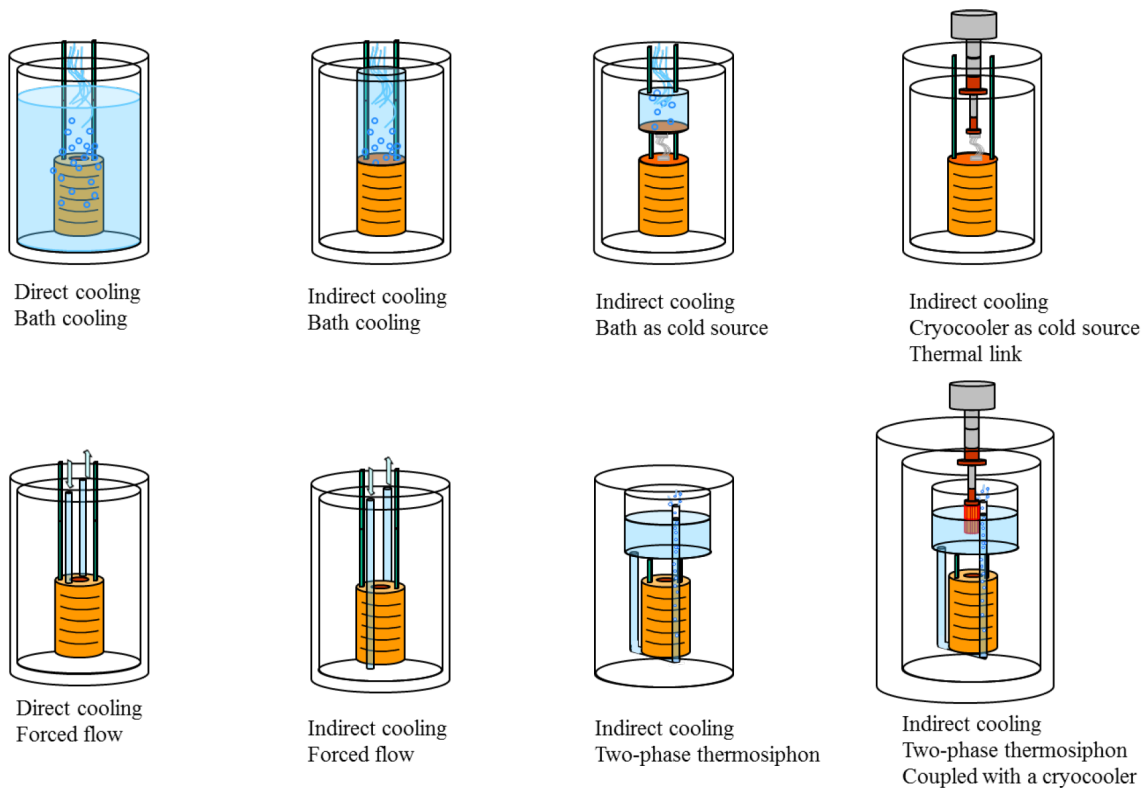

**Fig. 18:** Schematic examples of various cooling methods

The principal reason to use a 'wet system' (or direct contact) cooling method is to evacuate a large heat load or maintain a uniform temperature within the system. The direct contact between the cryogenic fluid and the system not only ensures a large heat transfer rate, but also allows the liquid to serve as an enthalpy reserve to overcome large transient heat perturbations. This can be achieved with a liquid bath, a stagnant coolant (He II pressurized or saturated), or a single-phase flow (supercritical). The first examples that come to mind are the accelerator magnets, although CICC magnets (ITER, 45 T NHMFL magnet), He II cooled magnets (Tore Supra, Iseult), and superconducting cavities are also cooled using the 'wet system' method (bath cooling).

## 3.2 Different methods of cooling

### 3.2.1 Baths

The most common cooling method is to use a liquid bath in which the system is immersed. The simplest of these methods is the saturated bath, where the fluid at the free surface is at saturation ($T \approx T_{sat}$). It is a direct or indirect cooling method with no net liquid mass flow and where the main heat transfer process is essentially due to latent heat of vaporization. The main advantages are the simplicity of the cryogenic design and operation, the high heat transfer due to nucleate boiling, and, therefore, an almost constant surface temperature. If there is sub-cooling due to a hydrostatic pressure head, as is depicted for the helium cooling bath in Fig. 19, then there is an extra $\Delta T_{sub}$ that can be used before boiling is reached. The disadvantage of this technique, on the other hand, is that a large quantity of cryogen has to be handled, particularly in case of a quench of a cryomagnetic system (risk of pressure rise). A limited range of temperature along the liquid–gas saturation curve is available (He, 4.2 K; $H_2$, 20.4 K; $N_2$, 77.3 K; etc.). If $q > q_{cr}$, then the heat transfer coefficient in film boiling is an order of magnitude smaller than that in a nucleate boiling case and non-uniform cooling can result due to film vapour formation.

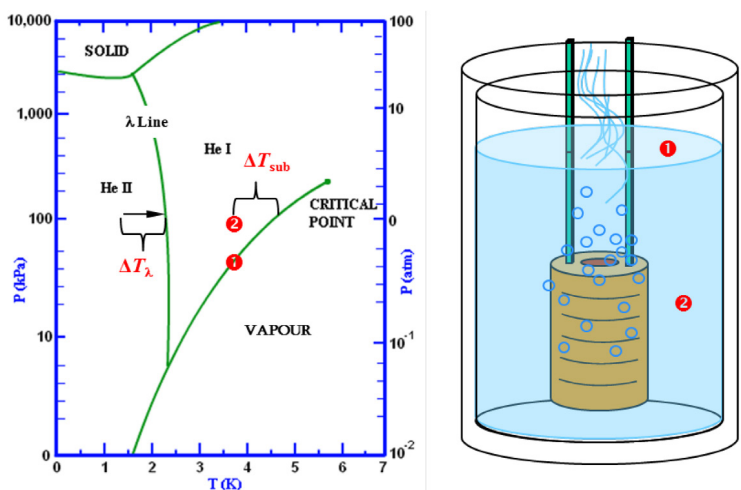

**Fig. 19:** An example of a helium cooling bath method

Superfluid helium (He II) bath cooling techniques are frequently employed to maintain a low temperature in superconducting accelerator cavities or certain accelerator magnets (below 2.17 K). The method exploits the extraordinary heat transfer capability of He II. The equivalent heat transfer conductivity is of the order of $k \approx 10^5$ W·m$^{-1}$·K$^{-1}$ and is independent of the flux, gravity, and surface orientation. The main thermal barrier of this cooling technique is the surface thermal resistance; the Kapitza resistance. For copper immersed in He II, the Kapitza resistance $R_k = 3 \times 10^{-4}$ K·m$^2$·W$^{-1}$ [23]; and for Kapton, $R_k = 10^{-3}$ K·m$^2$·W$^{-1}$ [24]. In the use of a saturated bath, the main risk is the handling of the volume below atmospheric pressure (leaks inducing air contamination). For pressurized baths most of the cryostat is above atmospheric pressure (as in the case of LHC magnets) which eliminates the risk of leaks. To take advantage of the cooling capacity of superfluid helium (i.e. stay in the superfluid state), the temperature superheat in the magnet has to be lower than $\Delta T_\lambda$ (see Fig. 19). Going lower in temperature is obviously costly for a large heat load (oversizing of pumping unit) and also complicates the design and operation. A detailed presentation of superfluid helium heat transfer is done in a different chapter of this volume.

### 3.2.2 Forced flow

Forced flow is one of the methods used to reduce the amount of cryogen, especially in indirect cooling composed of a network of peripheral tubes. Another advantage is the adjustable heat transfer rate with mass flow rate. An example of the forced flow of hydrogen is shown in Fig. 20 [13]. High heat transfer can be achieved, of the order of $10^4$ W·m$^{-2}$, in different cryogens in single phase. For a single-phase flow, the main disadvantages include: the pressurization system, or the implementation of the circulation pump and its maintenance at low temperature; the implementation of the heat exchanger system to sub-cool the fluid; and the temperature range limitation due to finite sub-cooling. For the example given in Fig. 20, the maximum allowable $\Delta T$ is 5 K prior to reaching the two-phase boiling regime (the onset of nucleate boiling).

Cooling in the supercritical state can be grouped in the single-phase category and one of its main advantages is a heat transfer coefficient comparable to that of pool boiling in helium. The heat transfer characteristics are $q \approx 10^4$ W·m$^{-2}$ for $\Delta T \approx 1$ K for helium [14]. But one needs to have a 'heavier' cryogenic installation to ensure a periodic re-cooling for operation (a series of large magnets) and to maintain a pressure above $P_{crit}$ (2.25 bars Absolute for He) in the system (usually $3 < P_{loop} < 6$ bars Absolute along the cooling circuit). Another advantage of cooling with supercritical fluid is the lack of hydraulic instabilities in the single-phase flow compared to a two-phase flow.

However, the use of two-phase forced flow cooling does have some advantages. The first is the guarantee of having an almost isothermal flow due to the high heat transfer, even at high vapour quality, in this flow regime. For the example of a 10 mm diameter horizontal tube in helium, the following cooling characteristics can be obtained: $q_{max} \approx 3 \times 10^3$ W·m$^{-2}$ for a mass flow rate of $6 \times 10^3$ kg·s$^{-1}$ and for $T \approx 1$ K with no initial sub-cooling [25]. The main drawbacks are the limited range of temperature cooling (LHe < 5.2 K, 14 < LH$_2$ < 20.4 K and 65 < LN$_2$ < 77.3 K, for example) and the non-uniform cooling if the vapour and the liquid are not homogeneous in their motion. Moreover, if the heat flux $q > q_{cr}$, then there is film boiling and the heat transfer rate becomes an order of magnitude smaller.

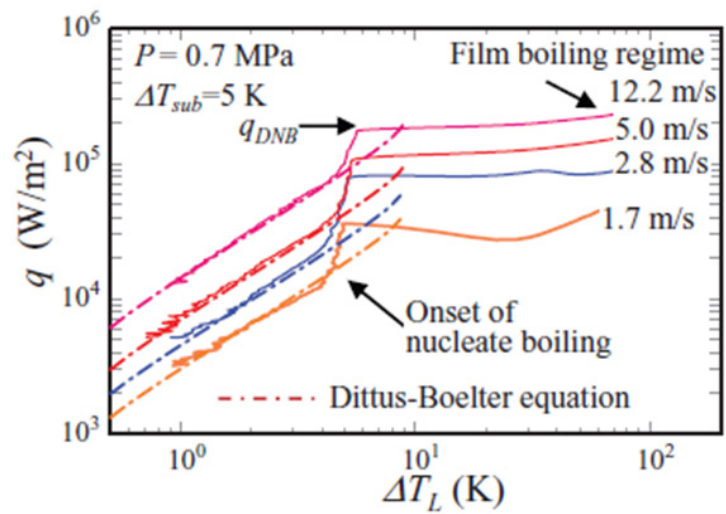

**Fig. 20:** The boiling curve in forced hydrogen flow, reprinted from [13] with permission of Elsevier

### 3.2.3 *Natural and two-phase circulation loops*

Circulation loops are an auto-tuned mass flow rate system in which the flow is created by the weight unbalance between the heated branch and the feeding branch of the loop due to vaporization or decreased vapour density, as described in Fig. 21. The main advantage is that there is no need for a circulating system (centrifugal pump, forced pressure differential, etc.). We can distinguish two types of circulation loops: the 'Open Loop', where the boil-off goes out of the system (as in Fig. 21) to be re-liquefied somewhere else before refilling the reservoir permanently to avoid a dry-out; and the 'Closed Loop', where the vapour is re-condensed in a closed reservoir with a heat exchanger.

The operating principle is the same if a single-phase fluid is used in the circulation loop, where density variation creates fluid motion. In nitrogen, vertical single-phase natural convection has been studied and the heat transfer coefficient as found in Fig. 21 is $q \approx 4 \text{ kW} \cdot \text{m}^{-2}$, with a mass flow rate of 40 g·s$^{-1}$ and $\Delta T \approx 3$ K for a Ø10 mm tube [26].

In a vertical configuration, two-phase flow circulation loops also have high heat transfer rates. In nitrogen, the heat flux is around $10^4$ W·m$^{-2}$ Ø10 mm tube for Ø10 mm and a mass flow rate of 40 g·s$^1$ for $\Delta T \approx 2.5$ K [25], whereas for helium the heat flux is around $10^3$ W·m$^{-2}$ for Ø10 mm and a mass flow rate of 20 g·s$^{-1}$, and $\Delta T \approx 0.3$ K [27]. When the heated section is horizontal, the heat transfer is as high as that for the vertical case, but there are flow instabilities at low heat flux flow [28]. The natural circulation loop in liquid helium can be modelled easily, with good accuracy, using the homogeneous flow model [29, 30].

### 3.2.4 *Cryogen-free cooling and the coupled system*

Today, cryocoolers can provide sufficient power to cool down and maintain a small cryomagnetic system at low temperature. In certain applications, one can use a conductive thermal link between the cold source (the cryocooler) and the low-temperature device. This is known as the cryogen-free cooling method. Even if the obvious advantages of easy implementation (no liquid, no heat exchanger, no transfer line, etc.) are very attractive, one has to be very accurate in the thermal design, since these cryocoolers provide a finite cooling power at a prescribed temperature. Therefore if the real power to be extracted exceeds the maximum power of the cryocooler, the working temperature will inevitably be higher than expected. The other technical issue to keep in mind is that a thermal link is necessary to distribute the cooling power over the entire system with this 'point-source' of cold that is the cryocooler. For a fully conductive thermal link, the thermal diffusion in the thermal link limits the cooling for transient events.

One of the solutions to overcome the disadvantages of the cryogen-free cooling method is to use a thermal link with a fluid. Several methods are under development and we will briefly present some of them.

Capillary-pumped devices have been used for many temperature ranges. The operating principle is identical whatever the temperature range. A flow is created by capillary pressure in a porous medium at the liquid/vapour interface. The simplest system is a heat pipe, with a wick inside the pipe serving as porous media. An typical example of such a system has been presented by Kwon [31]. This wick-based heat pipe, developed at the nitrogen temperature range, can transfer around 50 W between the cold source and the system to be cooled with a temperature difference of around 10 K. Here, the large temperature difference between the cold source and the warm source is the main disadvantage. More sophisticated cooling systems have been designed, especially for space applications; for example, the cryogenic loop heat pipe, which is basically a heat pipe in a loop configuration to create a mass flow rate. At nitrogen temperature, the most powerful ones can transmit 40 W for $\Delta T = 6$ K between the cold source and the system [32].

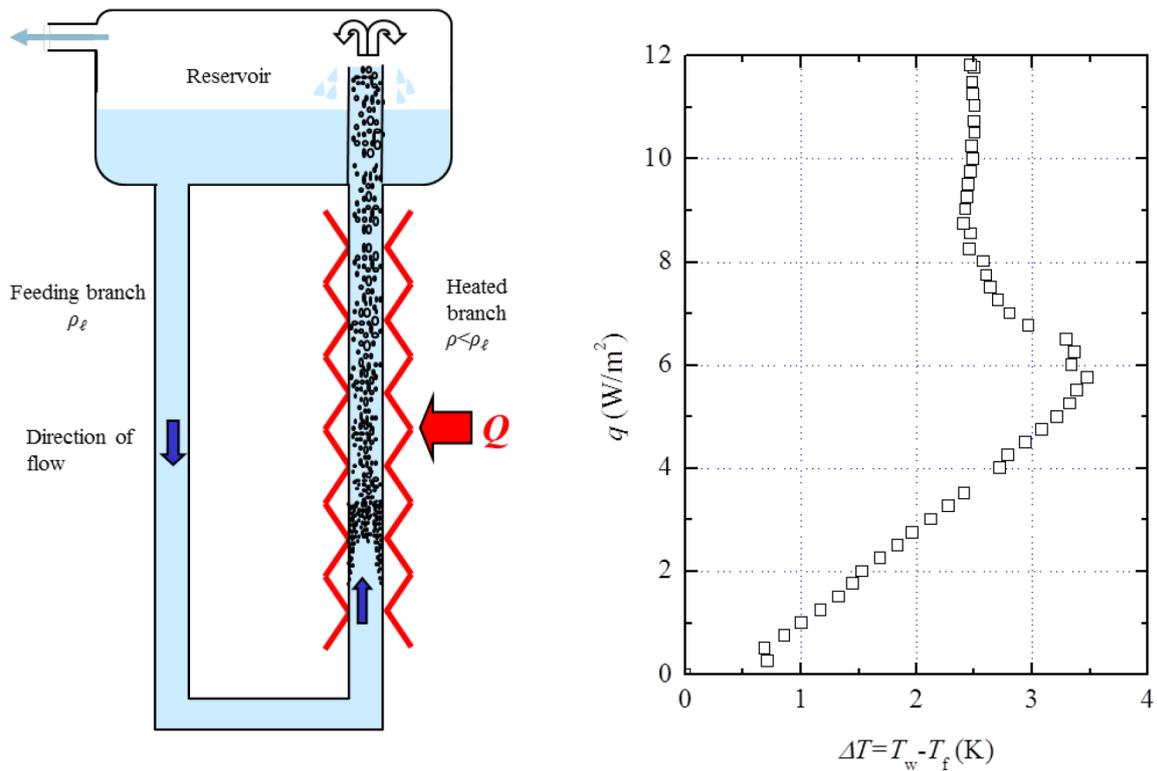

**Fig. 21:** A schematic of a vertical circulation loop: the boiling curve in a 1-m high nitrogen loop [26]

Another phase-change based thermal link is the Oscillating Heat Pipe (OHP). An OHP generally consists of a capillary tube, wound in a serpentine manner, connecting the ends to the inlets. The simplicity of its construction and the versatile geometry are the main advantages. The OHP utilizes the pressure change due to volume expansion and contraction during a phase transition to excite the oscillation of liquid slugs and vapour bubbles between the evaporator and the condenser. This system can be used with different fluids and exhibit large equivalent heat transfer conductivity (Table 7). As for the other capillary-pumped devices, the main drawback is the large temperature difference between the condenser (cooling part) and the evaporator (heating part), as shown in Table 8 [33].

The so-called 'vertical thermosyphon' is also an option to consider when creating a thermal link. The working principle is somewhat similar to the above-mentioned circulation loop and is based on the weight unbalance, but here there is a single vertically oriented tube. The liquid flows down the wall and the vapour flows in a counter-current manner to the liquid at the centre of the tube. One example in nitrogen can be found in [34]. This thermosyphon is able to transmit 20 W with a $\Delta T$ of 10 K at nitrogen temperature.

**Table 8:** The performance of an OHP for different fluids [33]

| Fluid | Heat input (W) | Temperature of cooling part (K) | Temperature of heating part (K) | $K_{eff}$ (kW·m$^{-1}$·K$^{-1}$) |
|---|---|---|---|---|
| $H_2$ | 0–1.2 | 17–18 | 19–27 | 0.5–3.5 |
| Ne | 0–1.5 | 26–27 | 28–34 | 1–8 |
| $N_2$ | 0–7 | 67–69 | 67–91 | 5–18 |

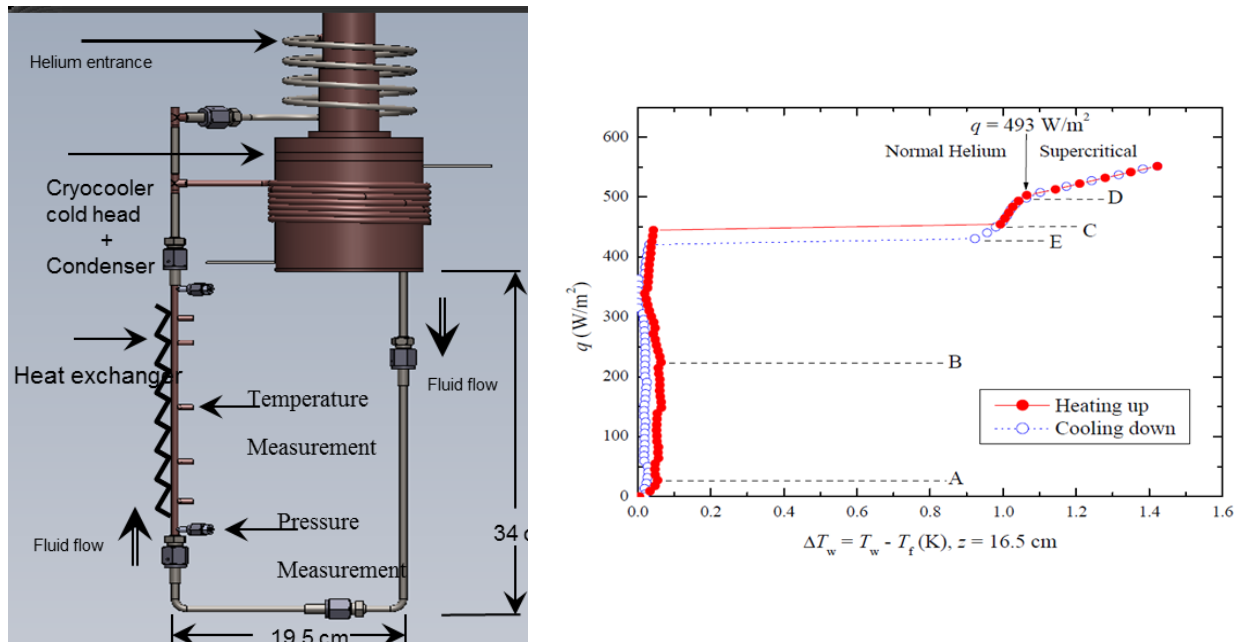

**Fig. 22:** (a) A schematic of a couped circulation loop with a cryocooler. (b) A typical boiling curve reprinted from [35] with permission of Elsevier.

Finally, a small natural circulation loop coupled with a cryocooler can be used to serve as a self-sustaining thermal link at cryogenic temperatures. The principle of this link is shown in Fig. 22. It is mainly composed of a condenser/phase separator cooled by the second stage of a cryocooler. Connected to the condenser, one can find a loop made of the heated branch (heat exchanger) and the feeding branch. Liquid helium flows through this loop and is re-condensed in the condenser. The heat transfer coefficient for a 4 mm diameter tube loop, in helium around 4.2 K, is 5000 W·m$^{-2}$·K$^{-1}$, with a critical heat flux of 500 W·m$^{-2}$ [35].

## References

[1] Cryocomp, Eckels Engineering, v. 3.06. Cryodata Inc., Florence, SC 29501, USA.

[2] J.W. Ekin, *Experimental Techniques for Low Temperature Measurements* (Oxford University Press, Oxford, 2006).

[3] R.F. Barron, *Cryogenic Heat Transfer* (Taylor & Francis, Philadelphia, PA, 1999).

[4] J.C. Johnson. A compendium of the properties of materials at low temperatures, part 1. WADD Tech. Rep. 60-56 (US Government Printing Office, Washington, DC, 1960).

[5] K.H. Hawks and W. Cottingham, Total normal emittances of some real surfaces at cryogenic temperatures, in *Advances in Cryogenic Engineering*, Ed. K. Timmerhaus (Plenum Press, New York, 1970), Vol. 16, pp. 467–474.

[6] W. Obert, Emissivity measurements of metallic surfaces used in cryogenic applications, in *Advances in Cryogenic Engineering,* Ed. R. Fast (Plenum Press, New York, 1982), Vol. 27, pp. 293–300.

[7] F.P. Incropera and D.P. DeWitt, *Fundamentals of Heat and Mass Transfer*, 5th ed. (John Wiley & Sons, New York, 2002).

[8] T. Nast. Radiation heat transfer, in *Handbook of Cryogenic Engineering*, Ed. J.G. Weisend (Taylor & Francis, Philadelphia, PA, 1998), p. 186.

[9] M.A. Hilal and R.W. Boom, *Int. J. Heat Mass Transfer* **23** (1980) 697–705.

[10] J.A. Clark, Cryogenic heat transfer, in *Advances in Heat Transfer*, Eds. T.F. Irvine and J.P. Hartnett (Academic Press, New York, 1968), pp. 325–517.

[11] D.E. Daney, *Int. J. Heat Mass Transfer* **19**(4) (1976) 431–441.

[12] Y. Shirai *et al.*, *Cryogenics* **50**(6–7) (2010) 410–416.

[13] H. Tatsumoto *et al.*, *Phys. Procedia* **36** (2012) 1360–1365.

[14] P.J. Giarrantano *et al.*, Forced convection heat transfer to subcritical helium I, in *Advances in Cryogenic Engineering*, Ed. K. Timmerhaus (Plenum Press, New York, 1974), Vol. 19, pp. 404–416.

[15] K. Ohira *et al.*, *Cryogenics* **51**(10) (2011) 563–574.

[16] R.V. Smith, *Cryogenics*. **9**(1) (1969) 11–19.

[17] M. Kida *et al.*, *J. Nucl. Sci. Technol.* **18**(7) (1981) 501–513.

[18] D.N. Lyon, Boiling heat transfer and peak nucleate boiling fluxes in saturated liquid helium bertween the λ and critical temperatures, in Advances in Cryogenic Engineering, Ed. K. Timmerhaus, (Plenum Press, New York, 1964), Vol. 10, pp. 371–379.

[19] Y.A. Kirichenko *et al.*, *Cryogenics* **23**(4) (1983) 209–211.

[20] J.C. Chen, *Ind. Engng Chem. Proc. Des. Dev.* **5**(3) (1966) 322–339.

[21] M.M. Shah, *ASHRAE Trans.* **82**(2) (1976) 66–86.

[22] H. Steiner and J. Taborek, *Heat Transfer Engng* **13**(2) (1992) 43–69.

[23] A. Iwamoto *et al.*, *Cryogenics* **41**(5–6) (2001) 367–371.

[24] B. Baudouy, *Cryogenics* **43**(12) (2003) 667–672.

[25] M. Mahé, Etude des propriétés d'échange thermique de l'hélium diphasique en convection forcée, Ph.D. thesis, Université Pierre et Marie Curie, Paris, 1991.

[26] B. Baudouy, *AIP Conf. Proc.* **1218** (2010) 1546–1553.

[27] L. Benkheira *et al.*, *Int. J. Heat Mass Transfer* **50**(17–18) (2007) 3534–3544.

[28] B. Gastineau *et al.*, *IEEE Trans. Appl. Supercond.* **22**(3) (2012) 900–1004.

[29] B. Baudouy, *AIP Conf. Proc.* **1434** (2012) 717–723.

[30] B. Baudouy *et al. Cryogenics* **53** (2013) 2–6.

[31] D.W. Kwon and R.J. Sedwick, *Cryogenics* **49** (2009) 514–523.

[32] Y. Zhao *et al.*, *Int. J. Heat Mass Transfer* **54** (2011) 3304–3308.

[33] K. Natsume *et al.*, *Cryogenics* **51** (2011) 309–314.

[34] A. Nakano *et al.*, *Cryogenics* **38**(12) (1998) 1259–1266.

[35] Y. Song *et al.*, *Int. J. Heat Mass Transfer* **66** (2013) 64–71.

**Bibliography**

**Journal and conference proceedings**

*Advances in Cryogenic Engineering*, Vols. 1–57, Proceedings of the Cryogenic Engineering and International Cryogenic Materials Conference (USA).

*Cryogenics*, Elsevier Science (http://www.journals.elsevier.com/cryogenics/).

Proceedings of the International Cryogenic Engineering Conference (Europe/Asia).

**Monographs**

R.F. Barron, *Cryogenic Heat Transfer* (Taylor & Francis, Philadelphia, PA, 1999).

J.W. Ekin, *Experimental Techniques for Low Temperature Measurements* (Oxford University Press, Oxford, 2006). http://www.oup.com

T.M. Flynn, *Cryogenic Engineering* (Marcel Dekker, New York, 1997).

W. Frost, *Heat Transfer at Low Temperature* (Plenum Press, New York, 1975).

S.W. Van Sciver, *Helium Cryogenics*, 2nd ed. (Springer, New York, 2012).

J.G. Weisend (Ed.), *Handbook of Cryogenic Engineering* (Taylor & Francis, Philadelphia, PA, 1998).

**Databases**

- Cryocomp, Eckels Engineering, v. 3.06. Cryodata Inc., Florence SC 29501, USA.
- Hepak, Gaspak, MetalPak, Cryodata Incorporated.
- NIST Database (http://cryogenics.nist.gov).